\documentclass[twocolumn,showpacs,preprintnumbers,prb,fleqn]{revtex4}
\usepackage{graphicx,amsfonts,amsbsy}

\begin{document}
%
%
\renewcommand{\Re}{\operatorname{Re}}
\renewcommand{\Im}{\operatorname{Im}}
\newcommand{\Tr}{\operatorname{Tr}}
\newcommand{\sign}{\operatorname{sign}}
\newcommand{\dd}{\text{d}}
\newcommand{\q}{\boldsymbol q}
\newcommand{\p}{\boldsymbol p}
\newcommand{\rr}{\boldsymbol r}
\newcommand{\pp}{p_v}
\newcommand{\vv}{\boldsymbol v}
\newcommand{\I}{{\rm i}}
\newcommand{\pphi}{\boldsymbol \phi}
\newcommand{\ds}{\displaystyle}
\newcommand{\be}{\begin{equation}}
\newcommand{\ee}{\end{equation}}
\newcommand{\bea}{\begin{eqnarray}}
\newcommand{\eea}{\end{eqnarray}}
\newcommand{\Acl}{{\cal A}}
\newcommand{\Rcl}{{\cal R}}
\newcommand{\Tcl}{{\cal T}}
\newcommand{\Tmin}{{T_{\rm min}}}
\newcommand{\Toff}{{\langle \delta T \rangle_{\rm off} }}
\newcommand{\Roff}{{\langle \delta R \rangle_{\rm off} }}
\newcommand{\RoffI}{{\langle \delta R_I \rangle_{\rm off} }}
\newcommand{\RoffII}{{\langle \delta R_{II} \rangle_{\rm off} }}
\newcommand{\dg}{{\langle \delta g \rangle_{\rm off} }}
\newcommand{\rd}{{\rm d}}
\newcommand{\br}{{\bf r}}
\newcommand{\la}{\langle}
\newcommand{\ra}{\rangle}
\newcommand{\ua}{\uparrow}
\newcommand{\da}{\downarrow}
\newcommand{\nn}{\nonumber}

%
%
\draft

\title{Spin interference effects in ring conductors subject to 
Rashba coupling}

\author{Diego Frustaglia}
\altaffiliation[Present address: ]{NEST-INFM \& Scuola Normale Superiore, 56126 Pisa, Italy.}
\affiliation{Institut f{\"u}r Theoretische Festk{\"o}rperphysik, Universit{\"a}t Karlsruhe, 76128 Karlsruhe, Germany}
\author{Klaus Richter}
\affiliation{Institut f{\"u}r Theoretische Physik, Universit{\"a}t Regensburg, 93040 Regensburg, Germany}



\date{\today}

%
\begin{abstract}

Quantum interference effects in rings provide suitable means for 
controlling spin at mesoscopic 
scales. Here we apply such control mechanisms to coherent 
spin-dependent transport 
in one- and two-dimensional rings subject to Rashba spin-orbit coupling.
We first study the spin-induced modulation of unpolarized currents
as a function of the Rashba coupling strength. The results suggest the 
possibility of all-electrical spintronic devices. Moreover, we find 
signatures of Berry phases in the conductance previously unnoticed.
Second, we show that the polarization direction of initially polarized,
transmitted spins can be tuned via an additional small magnetic 
control flux. In particular, this enables to precisely reverse the 
polarization direction at half a flux quantum. 
We present full numerical calculations for realistic 
two-dimensional ballistic microstructures and explain our
findings in a simple analytical model for one-dimensional rings.
\end{abstract}


\maketitle


\narrowtext
%
\section{Introduction}

In the last decade the field of {\it quantum electronics} \cite{S98,AAFKN98} 
has received extraordinary attention from both experimental and theoretical 
physics communities. Special effort has been made towards control and
engineering of the spin degree of freedom at the mesoscopic scale, 
usually referred to as {\it spintronics}. \cite{P98,WABDMRCT01} 
The major problem faced in this field 
is the generation of spin-polarized carriers and their 
appropriate manipulation in a controllable environment, preferable in 
semiconductors. 
Since the original proposal of the spin field effect transistor 
by Datta and Das, \cite{DD90} significant progress has been
made  \cite{W2002}  though the realization of a spin transistor 
still remains as a challenge.
Setups based on intrinsic spin-dependent properties
of semiconductors, as the Rashba effect \cite{R60,BR84} for a two-dimensional
electron gas (2DEG) confined to an asymmetric potential well,
appear to be of particular interest owing to the 
convenient means of all-electrical control through additional gate 
voltages.\cite{NATE97} 
In addition, coherent ring conductors enable to exploit the distinct
interference effects of electron spin {\em and} charge which arise
in these doubly connected geometries.  This opens up 
the area of spin-dependent Aharonov-Bohm physics, including
topics such as Berry phases, \cite{B84,bphase} 
spin-related conductance modulation,\cite{NMT99,MSC99} persistent currents,
\cite{LGB90,SGZ03} spin filters \cite{PFR03} and detectors,\cite{ID03} spin rotation,\cite{MSC02,CMC03}  and
spin switching mechanisms \cite{FHR01,FHR03,HSFR03}.


In this article we focus on two different aspects of spin-interference  
in ballistic one- and two-dimensional (1D and 2D) ring geometries subject 
to Rashba spin-orbit coupling.\cite{note14} 
First, motivated by the work of Nitta et al.,\cite{NMT99} in  Sec.~\ref{HRM}
we revisit the subject of spin-induced modulation of unpolarized 
currents using the Hamiltonian for 1D rings recently 
introduced by Meijer et al., \cite{MMK02} which sligthly differs from the one 
used previously.\cite{AL-G93,NMT99,CCRK97} Taking into account the
corresponding appropriate eigenstates, 
we derive in Sec.~\ref{RMSUC} the modulation profile of the 
conductance as a function of the Rashba coupling strength and
extract distinct effects due to the presence of Berry phases which
have not been recognized in earlier work.\cite{NMT99}
The 1D results are later compared with independent fully  numerical
calculations for 2D rings. The imprints of the 
Rashba coupling (strength) on the overall conductance 
is remarkable, pointing towards the possibility of 
all-electrical spintronics devices. 

Second, and motivated by our previous work 
on spin control in the presence of external 
inhomogeneous magnetic fields,\cite{FHR01} 
we study in Sec.~\ref{MSPCSS} the magneto conductance 
of initially spin-polarized carriers traversing a ring geometry 
with Rashba spin-orbit interaction. We demonstrate by means of
numerical calculations for 2D ring systems that the spin 
orientation of polarized carriers can be tuned and even reversed by means of
an additional small magnetic control field. This implies a spin-switching 
mechanism which is probably more convenient for experimental realizations 
than our previous proposal,\cite{FHR01} since the orignally 
suggested external inhomogeneous magnetic field is now replaced
by the intrinsic effective field due to the Rashba interaction.

After a short summary in  Sec.~\ref{CONCL} we present 
details of our analytical approach in an Appendix.

\section{Model and relevant parameters }
\label{HRM}
 
\subsection{Hamiltonian}

\begin{figure} 
\begin{center}
\includegraphics[width=9cm,angle=0]{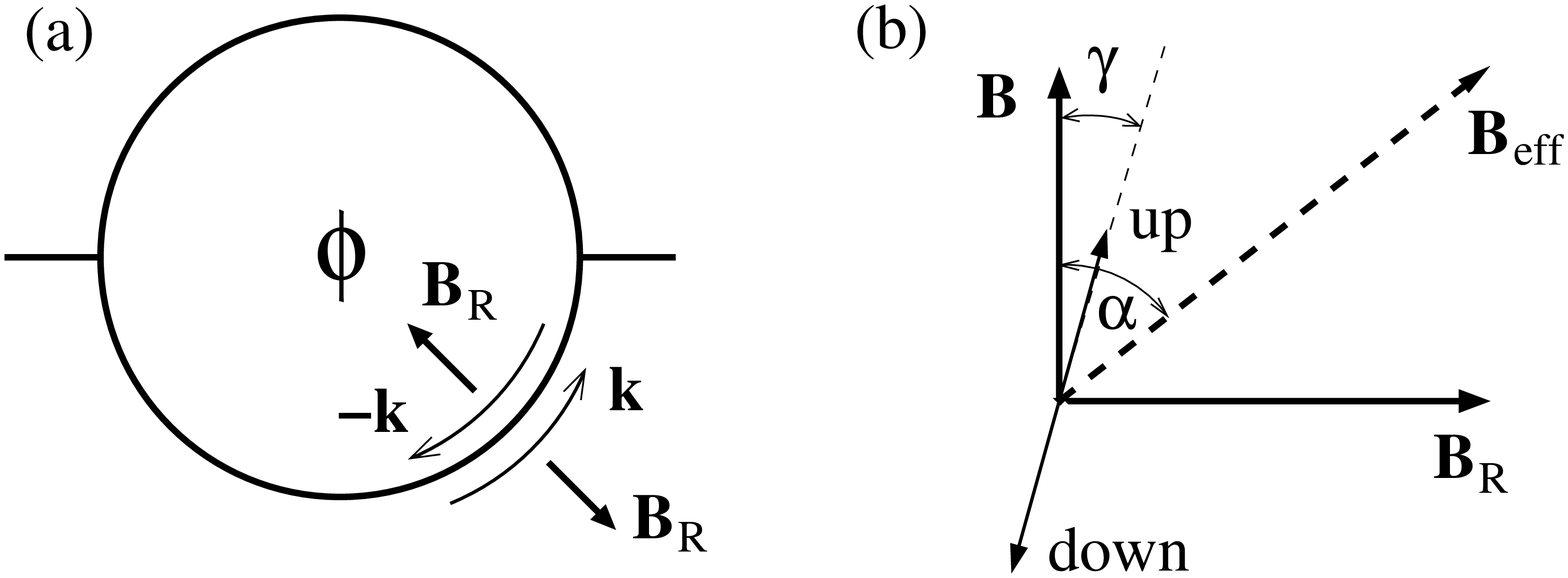}
\end{center}
\caption{
(a) 1D ring of radius $r_0$ subject to Rashba coupling in the presence of an 
additional, vertical magnetic field ${\bf B}$ (flux $\phi=\pi r_0^2 B$).
Spin carriers travelling around the ring see a momentum 
(${\bf k}$) dependent in-plane Rashba field ${\bf B}_{\rm R}$, which is 
orientationally inhomogeneous. (b) Up and down spin-eigenstates do not 
generally align 
with the total effective field ${\bf B}_{\rm eff}={\bf B}+{\bf B}_{\rm R}$.
}
\label{1Dring-rashba}
\end{figure}

The 2D quantum Hamiltonian for particles of charge $-e$ ($e>0$) and effective 
mass $m^*$ subject to Zeeman and Rashba coupling with coupling constants
$\mu$ and $\alpha_{\rm R}$, respectively, reads 
\bea
\label{H2D}
H_{\rm 2D}=\frac{1}{2 m^*} \boldsymbol{\Pi}^2+\mu {\bf B}\cdot 
\boldsymbol{\sigma}+ \frac{\alpha_{\rm R}}{\hbar}
\left(\boldsymbol{\sigma} \times \boldsymbol{\Pi}\right)_z + V({\bf r}) \; ,
\eea
where $\boldsymbol{\sigma}$ is the vector of the Pauli spin matrices,
$\boldsymbol{\Pi}={\bf p} + (e/c) {\bf A}$, and ${\bf B}=\nabla \times 
{\bf A}$. The electrostatic potential $V({\bf r})$ defines, e.g., the 
confining potential of a 2D ballistic conductor.
Recently it has been shown \cite{MMK02} that taking the limit from 2D to
1D rings (Fig.~\ref{1Dring-rashba}(a)) 
has to be performed by carefully considering in the 
above Hamiltonian (\ref{H2D}) the radial wave functions 
in the presence of a narrow confinement. As a consequence, the corresponding 
1D Hamiltonian for 
a ring of radius $r_0$ in the presence of a vertical magnetic field 
${\bf B}=(0,0,B)$ reads\cite{MMK02,note1}
\bea
H_{\rm 1D}&=&\frac {\hbar \omega_0}{2} \left( -i \frac{\partial}{\partial 
\varphi} + \frac{\phi}{\phi_0} \right)^2 + \frac {\hbar \omega_B}{2} \sigma_z 
\nn \\
&+& \frac {\hbar \omega_{\rm R}}{2} (\cos\varphi~\sigma_x + \sin\varphi~\sigma_y)\left( -i \frac{\partial}{\partial 
\varphi} + \frac{\phi}{\phi_0} \right) \nn \\
&-&i \frac {\hbar \omega_{\rm R}}{4} (\cos\varphi~\sigma_y - 
\sin\varphi~\sigma_x),
\label{H1D}
\eea
where we have introduced the polar angle $\varphi$, the frequencies 
$\omega_0=\hbar/(m^*r_0^2)$, $\omega_B=2\mu B/\hbar$ and 
$\omega_{\rm R}=2\alpha_{\rm R}/(\hbar r_0)$, and the magnetic fluxes 
$\phi=\pi r_0^2 B$ and $\phi_0=hc/e$. 

The 1D eigenstates of (\ref{H1D}) have the general form
\bea
\label{spinor}
\Psi_{\lambda,n}^s(\varphi)=\exp(i \lambda n\varphi)~\chi_{\lambda,n}^s \quad ; \quad
\chi_{\lambda,n}^s = \left( \begin{array}{c}
\chi_1 \\
\chi_2~e^{i \varphi}
\end{array}
\right) \; .
\eea
Here, the spin components $\chi_{1,2}$ depend in principle on the
travel direction $\lambda=\pm 1$, orbital quantum number $n \ge 0$ ($n$ integer), 
and spin $s=\pm 1$.  The spin carriers being subject to 
 $H_{\rm 1D}$ experience an effective magnetic 
field ${\bf B}_{\rm eff}={\bf B}+{\bf B}_{\rm R}$ composed of the
external field $\bf B$ and the momentum dependent field ${\bf B}_{\rm R}$ 
arising from the Rashba coupling. ${\bf B}_{\rm R}$ lies in 
the plane of the ring.
${\bf B}_{\rm eff}$ encloses a tilt angle $\alpha$ with the 
$z$-axis given by $\tan \alpha=B_{\rm R}/B=\omega_{\rm R} 
(n'+1/2)/\omega_B$ with $n'=\lambda n+\phi/\phi_0$ (see 
Appendix~\ref{App1} for further details). The exact orientation of 
${\bf B}_{\rm R}$ is determined by the magnitude and sign of the momentum, 
namely $\lambda n$, i.e.\
 spins travelling in opposite directions are subject to a different 
${\bf B}_{\rm R}$.
Moreover, Eq.~(\ref{H1D}) implies that the orientation of 
${\bf B}_{\rm eff}$ varies spatially. \cite{note2} This means that, in 
general, the corresponding spin eigenstates (\ref{spinor})
are {\it not aligned} with ${\bf B}_{\rm eff}$ 
(see Fig.~\ref{1Dring-rashba}(b)). On the 
contrary, they are characterized by a different tilt angle $\gamma$ determined 
by the relative magnitude of the spinor components $\chi_1$ and $\chi_2$. 
However, in the 
limit of strong spin-orbit coupling, the so-called {\it adiabatic} regime, 
the spin eigenstates follow the local direction of the effective field, and 
$\gamma \rightarrow \alpha$ (leading to Berry phases\cite{B84}). This limit is 
reached if
the adiabaticity parameter $Q=Q_{B}+Q_{\rm R}$ satisfies $Q \gg 1$, 
\cite{S92,AL-G93} where we have defined 
$Q_B=\omega_B/(\omega_0|n'+1/2|)$ and  $Q_{\rm R}$, particularly
relevant here, as
\begin{equation}
\label{eq:QR}
Q_{\rm R}=\omega_{\rm R}/\omega_0 \; .
\end{equation}
Hence, the adiabatic limit corresponds to the situation where a spin 
precesses many times during a full travel around the ring.

\subsection{1D eigenstates in the absence of an external magnetic field} 
\label{1Deigenstates}

For ${\bf B}= 0$ we have $\omega_{\rm B}=0$ and $\phi=0$ in Eq.\
(\ref{H1D}), and the Hamiltonian $H_{\rm 1D}$ simplifies considerably.
The resulting effective field reduces to the in-plane field
${\bf B}_{\rm eff}={\bf B}_{\rm R}$ with tilt angle $\alpha=\pi/2$. 
In this situation, the 1D eigenstates (\ref{spinor}) take the simple 
form (see Appendix \ref{App1} for details)
\bea
\label{spinor1}
\Psi_{+,n}^\ua(\varphi)&=&\exp(i n\varphi) \left( \begin{array}{c}
\sin \gamma/2 \\
\cos \gamma/2~e^{i \varphi}
\end{array}
\right), \\
\nn \\
\label{spinor2}
\Psi_{+,n}^\da(\varphi)&=&\exp(i n\varphi) \left( \begin{array}{c}
\cos \gamma/2 \\
-\sin \gamma/2~e^{i \varphi}
\end{array}
\right), \\
\nn \\
\label{spinor3}
\Psi_{-,n}^\ua(\varphi)&=&\exp(-i n\varphi) \left( \begin{array}{c}
\cos \gamma/2 \\
-\sin \gamma/2~e^{i \varphi}
\end{array}
\right), \\
\nn \\
\label{spinor4}
\Psi_{-,n}^\da(\varphi)&=&\exp(-i n\varphi) \left( \begin{array}{c}
\sin \gamma/2 \\
\cos \gamma/2~e^{i \varphi}
\end{array}
\right).
\eea
The corresponding tilt angle $\gamma$ is given by 
$\tan \gamma =Q_{\rm R}$, satisfying $\gamma \rightarrow \alpha=\pi/2$ in the 
adiabatic limit $Q_{\rm R} \rightarrow \infty$. 
Hence, we note that the spinors $\chi_{\lambda,n}^s$ in (\ref{spinor1})-(\ref{spinor4}) do not 
actually depend on $n$. 
Moreover, the associated
eigenenergies read
\be
\label{ES1}
E_{\lambda,n}^s = \frac{\hbar \omega_0}{2}\! \left[\left(\lambda n
+ \frac{1}{2}\right)^2 \! + \! \frac{1}{4} \! + \!
  s \left|\lambda n+ \frac{1}{2}\right|
\sqrt{1+Q_{\rm R}^2} \right] .
\ee
The above spin eigenstates (\ref{spinor1})-(\ref{spinor4}) are defined 
in such a way that the 
eigenenergies
(\ref{ES1}) are maximum 
for spin-up states. We will make use of these results in the following section 
for the study of transport properties.

\section{Rashba modulation of unpolarized currents}
\label{RMSUC}

We first consider the case where the 1D ring of 
Sec.~\ref{1Deigenstates} 
is symmetrically coupled to two contact leads 
(Fig.~\ref{1Dring-rashba}(a)) in order to
study the transport properties of the system subject to a constant, low bias voltage 
(linear regime).
To this end we calculate the zero temperature conductance $G$ based on the 
Landauer formula \cite{note9} 
\bea
\label{G}
G=\frac{e^2}{h} \sum_{m',m=1}^M \sum_{\sigma',\sigma} 
T_{m' m}^{\sigma' \sigma} \; ,
\eea 
where $T_{m' m}^{\sigma' \sigma}$ denotes the quantum probability of transmission
between incoming ($m,\sigma$) and outgoing ($m',\sigma'$) asymptotic states defined on 
semi-infinite ballistic leads. The labels $m,m'$ and $\sigma,\sigma'$ refer to the 
corresponding mode and spin quantum numbers, respectively. 
For 1D rings ($M=1$ in (\ref{G})) the transmission coefficients can be 
approximated to first order as follows: In the presence of Rashba coupling 
the energy splitting is such that particles with Fermi energy 
$E_{\rm F}$ can traverse the ring with four 
different wave numbers $n_\lambda^s$, depending on spin ($s$) and 
direction of motion ($\lambda$). The quantities $n_\lambda^s$ are obtained by solving
$E_{\lambda,n}^s=E_{\rm F}$ in Eq.~(\ref{ES1}) and do not require to be integer. 
Moreover, in this simple approach we assume perfect coupling between leads and ring
(i.e. fully transparent contacts), 
neglecting backscattering effects leading to resonances.   
Thus, incoming spins $|\sigma \rangle$ entering the ring at $\varphi=0$ propagate 
coherently along the four available channels and interfere at $\varphi=\pi$, 
leaving the ring in a mixed spin state $|\sigma_{\rm out} 
\rangle=\sum_{\lambda,s} \langle \chi_\lambda^s(0)|\sigma \rangle 
\exp(n_\lambda^s \pi) |\chi_\lambda^s(\pi)\rangle$.\cite{note3} 
Choosing a complete basis 
of incoming and outgoing spin states, the spin-resolved transmission 
probabilities are obtained as 
$T^{\sigma' \sigma}=|\langle \sigma' | \sigma_{\rm out} \rangle|^2$.  
After summation over the spin indeces $\sigma$ and $\sigma'$, we obtain
for the total conductance 
\bea
\label{1DGb}
G=\frac{e^2}{h}\left[1 \! + \!
\frac{1}{2}\left[\cos\pi(n_-^\da\!-\!n_+^\ua)+\cos\pi(n_-^\ua\!-\!n_+^\da)\right]\right].~~
\eea
Note that the phase difference acquired by opposite spin 
states travelling in opposite directions plays an important role for the 
modulation of the conductance.\cite{note4} The 
spin-dependent phases are signatures of the Aharonov-Casher effect \cite{AC84} for 
spins travelling in the presence of an electric field, which is the 
electromagnetic dual of the Aharonov-Bohm effect.  

By imposing $E_{\lambda,n}^s=E_{\rm F}$ in Eq.~(\ref{ES1}) we 
obtain
\bea
(n_-^\da-n_+^\ua)&=& 1+\sqrt{1+Q_{\rm R}^2} \; , \\
(n_-^\ua-n_+^\da)&=& 1-\sqrt{1+Q_{\rm R}^2} \; .
\eea
Inserting the above expressions into Eq.~(\ref{1DGb}) one finds 
the total conductance as a function of the dimensionless Rashba 
coupling strength $Q_{\rm R}$:
\begin{eqnarray}
\label{1DGc}
G & = & \frac{e^2}{h}\left\{1+ 
\cos\left[\pi\left(\sqrt{1+Q_{\rm R}^2}-1\right)\right]\right\} \\
& = & 
 \label{1DGphase}
 \frac{e^2}{h}\left\{1+ 
 \cos\left[\pi  Q_{\rm R} \sin \gamma -\pi (1-\cos \gamma)
 \right]\right\} \; ,
\end{eqnarray}
where we used $\tan \gamma =Q_{\rm R}$,
$\cos \gamma=1/\sqrt{1+Q_{\rm R}^2}$, and
$\sin \gamma=Q_{\rm R}/\sqrt{1+Q_{\rm R}^2}$.
Comparing Eq.~(\ref{1DGc}) with the corresponding result
 of Nitta, Meijer, and Takayanagi, \cite{NMT99}
\bea
\label{GNMT}
G_{\rm NMT}=\frac{e^2}{h}\left[1+ \cos\left(\pi Q_{\rm R}\right)\right] \; ,
\eea  
we recognize two main contributions to the phase in (\ref{1DGphase}): One is 
the 
Rashba phase $\varphi_{\rm R}=\pi Q_{\rm R} \sin \gamma$. This is 
similar to the phase $\pi Q_{\rm R}$\cite{note5} appearing in $G_{\rm NMT}$, 
Eq.~(\ref{GNMT}), but corrected 
by a factor $\sin \gamma$ accounting for the fact that the spinors are 
generally not
aligned with ${\bf B}_{\rm eff}$. In the limit of adiabatic spin 
transport both phases coincide (since $\sin \gamma \rightarrow 1$ as 
$Q_{\rm R}\rightarrow \infty$). 
Moreover, we find an additional Aharonov-Anandan phase \cite{AA87} 
contribution $\varphi_{\rm AA}=\pi(1-\cos \gamma)$ to (\ref{1DGphase}) 
absent in $G_{\rm NMT}$ and
related to the solid angle accumulated by the change of spinor orientation 
during transport. In 
the adiabatic limit, $\varphi_{\rm AA}$ tends to the corresponding Berry phase
$\varphi_{\rm B}=\pi(1-\cos \alpha)$ as 
$\cos \gamma \rightarrow \cos \alpha$ (where $\cos \alpha=0$, i.e.\ $\varphi_{\rm B}
= \pi$ in the present case). 

\begin{figure} 
\begin{center}
\includegraphics[width=4cm,angle=-90]{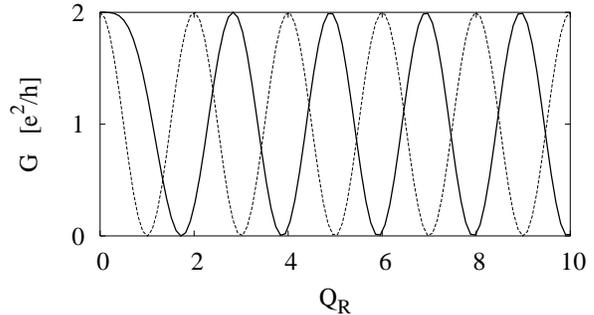}
\end{center}
\caption{
Conductance modulation profile of 1D rings (Fig.~\ref{1Dring-rashba}(a)) as a 
function of the dimensionless
Rashba strength $Q_{\rm R}$ in the absence of external magnetic field 
(${\bf B}=0$). The curves show our result (\ref{1DGc}) for
$G$ (solid line) compared to the originally incomplete $G_{\rm NMT}$ of 
Eq.~(\ref{GNMT}) (dashed line). 
}
\label{G-Gnmt}
\end{figure}

In Fig.~\ref{G-Gnmt} we plot for comparison our result, Eq.~ (\ref{1DGc}),
for $G$ together with $G_{\rm NMT}$, Eq.~(\ref{GNMT}), as a function
of the Rashba strength $Q_{\rm R}$. There we observe that while
$G_{\rm NMT}$ (dashed line) shows regular oscillations of period 2 in 
$Q_{\rm R}$-units, our result (solid line) exhibits quasi-periodic oscillations 
of period larger than 2 reflecting the fact that non-adiabatic spin 
transport ($\sin \gamma < 1$) takes place for small $Q_{\rm R}$. For 
$Q_{\rm R} \gg 1$ the period is tending to 2 as the adiabatic limit is approached.
In addition, a relative phase shift of magnitude $\pi$ survives between $G$ and
$G_{\rm NMT}$ for large $Q_{\rm R}$, coinciding with the appearance of the 
Berry phase $\varphi_{\rm B}=\pi$. As a consequence, minima in $G$ are 
obtained for even integers of $\sqrt{Q_{\rm R}^2 + 1}$, i.e.\
$Q_{\rm R}=\sqrt{3},\sqrt{15},\ldots$. These 
minima are reminiscent of those found for the conductance of rings 
subject to Zeeman spin-coupling to in-plane circular magnetic fields 
(instead of Rashba coupling) as a function of the corresponding adiabaticity 
parameter. \cite{FHR01} Moreover, we note that Eq.~(\ref{GNMT}),
predicting uniform oscillations as a function of the coupling strength, 
actually corresponds to the conductance of a 1D ring subject to 
a radial electric field of constant magnitude (instead of a 
vertical one as in the case of Rashba coupling).
\cite{CCRK97,MS92}
    
\begin{figure}
\begin{center}
\includegraphics[width=6cm,angle=0]{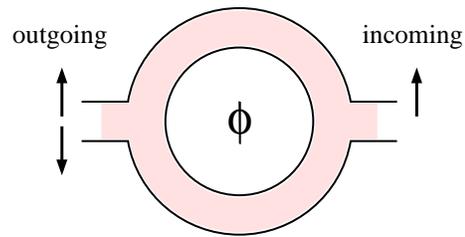}
\end{center}
\caption{
2D ring of mean radius $r_0$ and width $w$ used for numerical calculations 
of the conductance. The grey zone corresponds to the region subject to 
a finite Rashba coupling. This is switched on and off adiabatically 
within the leads by using a linear function. An additional, vertical magnetic 
field ${\bf B}$ generates a flux $\phi=\pi r_0^2 B$.
}
\label{2Dring-rashba}
\end{figure}

\begin{figure}[ht]
\begin{center}
\includegraphics[width=4cm,angle=-90]{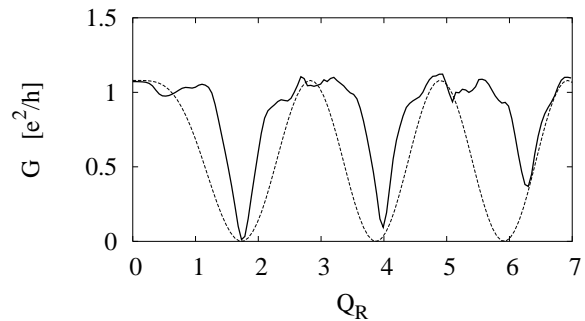}
\end{center}
\caption{
Numerical calculation of the conductance modulation profile (solid line) of 
a single-mode 2D ring (Fig.~\ref{2Dring-rashba}, aspect 
ratio $w/r_0\approx 0.3$) as a function of the dimensionless
Rashba strength $Q_{\rm R}$ in the absence of an external magnetic field 
(${\bf B}=0$). Dashed line: corresponding 1D result,
Eq.~(\ref{1DGc}) (same as solid line in Fig.~\ref{G-Gnmt}) including a 
fitting prefactor at $Q_{\rm R} = 0$ for comparison.
}
\label{Gnum-sym.ring}
\end{figure}

To complete the above discussion we present in the following the results of 
independent numerical calculations corresponding to more realistic 2D 
ring structures (Fig.~\ref{2Dring-rashba}). To this end we calculate the 
zero temperature 
conductance $G$ based on the Landauer formula (\ref{G}) by using a 
spin-dependent, recursive Green function technique\cite{note6} applied to the
2D Hamiltonian (\ref{H2D}). Unless otherwise stated, our numerical calculations 
correspond to a quantum transmission averaged within a small energy window\cite{note7} 
in order to smooth 
out energy-dependent oscillations related to resonances in the ring structure.
Moreover, the Rashba coupling is switched on and off adiabatically\cite{note12} within the 
leads by using a linear function.\cite{note13} 
Fig.~\ref{Gnum-sym.ring} (solid line) shows the result for a single-mode ring 
of mean radius $r_0$ and width $w$ 
(aspect ratio $w/r_0 \approx 0.3$) 
symmetrically coupled to two leads of the same width 
(see Fig.~\ref{2Dring-rashba}).  
This result is to be compared with that for the strictly 1D ring of 
Eq.~(\ref{1DGc}) (dashed line; overall scaling factor included).
Both curves present similar features on the whole. We observe that the 
first minimum of 
$G$ in Fig.~\ref{Gnum-sym.ring} coincides for both 1D and 2D calculations. 
However, as $Q_{\rm R}$ increases the 2D minima (solid line) undergo a small 
relative shift with respect to the 1D result (dashed line) and get less 
pronounced. This can be related to the finite aspect ratio of 
the ring: The strength $Q_{\rm R}$ can actually be written as 
$Q_{\rm R}=(r_0/w)\delta_{\rm R}$, where $\delta_{\rm R}=
\alpha_{\rm R} (2 m^*/\hbar^2) w$ is the parameter defining the strength 
of the Rashba coupling in 2D conducting wires of width $w$.\cite{DD90,MK01}
The weak coupling regime characterized by spin subband separation is defined 
for $\delta_{\rm R} \ll 1$. For the case of 1D wires and rings,
this condition is always
satisfied since $w=0$. For finite width (represented by 
the finite $w/r_0$ in our case) the situation is different, as we verify in
our results of Fig.~\ref{Gnum-sym.ring}. There, the first minimum at 
$Q_{\rm R}=\sqrt{3}$ (fitting  the above 1D result) corresponds 
to a relatively small coupling strength $\delta_{\rm R}\approx 0.5$. However, 
at the second minimum, $Q_{\rm R}\approx 4 > \sqrt{15}$, we 
already enter the strong coupling regime with $\delta_{\rm R} \approx 1.2$. 
As a consequence deviations from the 1D case in the corresponding 
conductance modulation profile arise. This tendency is less pronounced as 
$w/r_0 \rightarrow 0$ and the parameter $\delta_{\rm R}$ looses relevance. 

Moreover, we note that in Fig.~\ref{Gnum-sym.ring} the conductance minima 
of finite width rings (solid line) suffer the shift to {\it larger} 
values of $Q_{\rm R}$ 
with respect to the 1D results (dashed line) as $Q_{\rm R}$ 
increases. This suggests that the radial motion in 2D rings 
obstructs the approach to the regime of adiabatic spin 
transport, 
since a relatively larger coupling $Q_{\rm R}$ would be necessary 
for obtaining the same spinor tilt angle $\gamma$ according to the structure 
of the phase (\ref{1DGphase}).

Additionally, numerical results\cite{FRunp} not presented here indicate that 
the conductance of 2D ring structures supporting several open channels 
shows a modulation pattern similar to that of 
Fig.~\ref{Gnum-sym.ring}, provided that (i) the incoming and 
outgoing leads 
support just one open channel and (ii) the corresponding aspect 
ratio is small ($w/r_0 \ll 1$). 
Furthermore, ring structures of irregular shape (leading to ballistic 
backscattering enhancement) exhibit a halfing of the period in 
$G(Q_{\rm R})$ modulation profiles when compared with that 
of Fig.~\ref{Gnum-sym.ring}, similarly to what is predicted for disordered 
systems.\cite{MS92,MGE-W89,OE-W92}

\section{magnetoconductance of spin polarized currents and spin switching}
\label{MSPCSS}

In this section we discuss the possibility of controlling the spin 
orientation of {\it spin-polarized} carriers by means of distinct
interference effects in mesoscopic ring structures due to 
(charge and spin) quantum coherence.
Motivated by our previous work on spin-switching in the presence of
in-plane circular magnetic fields\cite{FHR01} we study here the 
magnetoconductance of incoming spin-polarized carriers,\cite{note10} now subject to 
Rashba interaction. 

\begin{figure} 
\begin{center}
\includegraphics[width=8cm,angle=0]{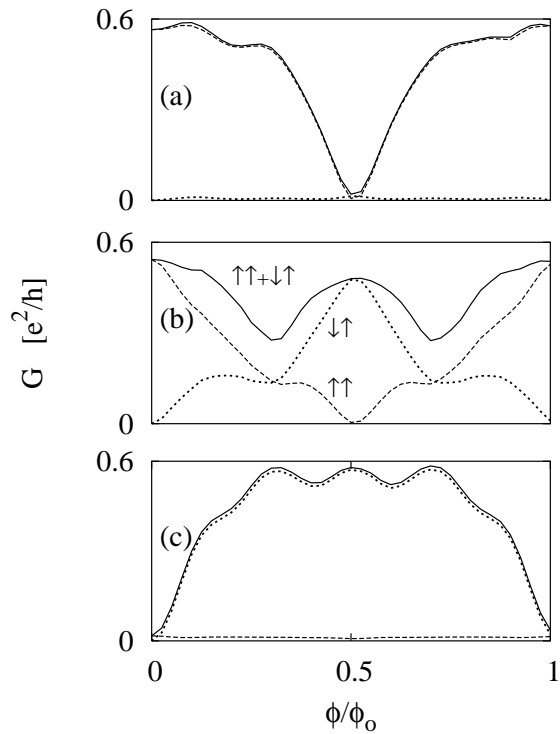}
\end{center}
\caption{
Numerical results for the conductance of spin-up polarized incoming carriers 
(see Fig.~\ref{2Dring-rashba}) through a single-mode 2D ring (aspect ratio 
$w/r_0\approx 0.3$) as a function of a flux $\phi=\pi r_0^2 B$ in the 
presence of Rashba coupling of increasing strength: $Q_{\rm R} \approx$ 
0.2 (a), 1.0 (b), and 1.7 (c) (see Fig.~\ref{Gnum-sym.ring} for comparison 
at $\phi=0$). The overall conductance (solid line) is split into its 
components $G^{\ua \ua}$ (dashed) and $G^{\da \ua}$ (dotted). 
Note the continuous change of the spin polarization with $\phi$ and the spin
switching at $\phi=\phi_0/2$. 
}
\label{spin-switch}
\end{figure}

The setup proposed is that of Fig.~\ref{2Dring-rashba}, representing
a 2D ring (aspect ratio $w/r_0\approx 0.3$) subject to Rashba coupling 
symmetrically coupled
to two leads. In addition, a weak magnetic field ${\bf B}$ is 
applied along the vertical axis leading to a flux $\phi$. 
Incoming and outgoing spin states are defined along the vertical axis as shown 
in Fig.~\ref{2Dring-rashba}. We consider spin-up polarized incoming 
particles\cite{note11} 
(equivalent results are obtained for spin-down incoming states).
Using the recursive Green function technique introduced in Sec.~\ref{RMSUC} 
we calculate numerically the spin-resolved conductances $G^{\ua \ua}$ and 
$G^{\da \ua}$, corresponding to outgoing spin-up and -down channels, 
respectively (see Fig.~\ref{2Dring-rashba}). 
In order to smooth out energy-dependent oscillations, the present numerical 
calculations correspond to an energy-averaged quantum transmission in a 
small energy window.\cite{note7} 
Our main results for a single-mode ring 
are summarized in Fig.~\ref{spin-switch}, showing the overall conductance
(solid line) split into its components
$G^{\ua \ua}$ (dashed line) and  $G^{\da \ua}$ (dotted line) as a function of
the magnetic flux $\phi$ for three different scaled Rashba strengths 
$Q_{\rm R}\approx$ 0.2, 1.0 and 1.7.
In the weak coupling limit, 
Fig.~\ref{spin-switch}(a), the overall conductance (solid line) shows the 
usual AB oscillations of period $\phi_0$ and is dominated by $G^{\ua \ua}$ 
(dashed line). As expected for weak spin-coupling, the spin polarization is 
almost conserved during transport.

More interesting features appear for the case of moderate coupling 
depicted in panel (b). There, both components $G^{\ua \ua}$ (dashed line)  
and $G^{\da \ua}$ (dotted line) contribute similarly to the overall 
conductance (solid line). However, the spin-polarization of the transmitted 
carriers
changes continuously as a function of the magnetic flux $\phi$: We note that
$G^{\da \ua}=0$ at $\phi=0$, while $G^{\ua \ua}=0$ at $\phi=\phi_0/2$. 
Hence, for zero flux all transmitted carriers conserve their original 
(incoming) spin-orientation, while for $\phi=\phi_0/2$ the transmitted 
particles reverse their spin-polarization. That means that by tuning the 
magnetic flux from $0$ to $\phi_0/2$ we can reverse the spin-polarization of 
transmitted particles in a controlled way. The setup of 
Fig.~\ref{2Dring-rashba} (i.e. AB ring subject to Rashba coupling) acts 
as a tunable spin-switch, similarly to our previous proposal for AB rings 
subject to inhomogeneous magnetic fields\cite{FHR01} with the advantage that 
in the present system the spin-dependent (Rashba) coupling can be 
electrically controlled.\cite{NATE97}
Moreover, such spin-switching mechanism 
is independent of the strength $Q_{\rm R}$, which determines only the 
amplitude of the spin-reversed current (see below). 

In addition to the above results we present in Fig.~\ref{spin-switch}(c)
calculations for a little larger strength $Q_{\rm R}\approx 1.7$, 
corresponding to 
the vicinity of the first minimum in Fig.~\ref{Gnum-sym.ring} for zero flux. 
There we see that the AB oscillations in the overall conductance (solid line) 
suffer a shift of $\phi_0/2$ with respect to the weak coupling case of panel 
(a). This is due to the additional phase of order $\pi$ acquired by 
the carriers for $Q_{\rm R} \approx \sqrt{3}$ (see Eq.~(\ref{1DGc}) and 
related paragraphs). Moreover, the overall conductance is dominated by the 
spin-reversed component $G^{\da \ua}$ (dotted line), while the complementary 
$G^{\ua \ua}$ (dashed line) is suppressed due to quantum interference.

As the coupling strength $Q_{\rm R}$ increases, we obtain a sequence of 
magnetoconductance profiles which reproduce periodically the different panels
of Fig.~\ref{spin-switch}, following the order (a)$\rightarrow$(b)$\rightarrow$(c)$\rightarrow$(b)$\rightarrow$(a)$\rightarrow$(b)$\rightarrow$(c)\ldots. 
Such periodical feature is related to the unbounded accumulation of the 
Rashba phase in (\ref{1DGphase}) as a function of $Q_{\rm R}$. As a 
consequence, Fig.~\ref{spin-switch}(a) is related to values of $Q_{\rm R}$ 
corresponding to maxima of the conductance in Fig.~\ref{Gnum-sym.ring},
while Fig.~\ref{spin-switch}(c) is associated with the vicinity of the 
minima in Fig.~\ref{Gnum-sym.ring}. Fig.~\ref{spin-switch}(b), 
where the spin-switching effect appears most clearly,
corresponds to intermediate values of $Q_{\rm R}$ lying between maxima and
minima of the conductance in Fig.~\ref{Gnum-sym.ring}.

We point out that this mechanism for reversing the spin polarization does not 
rely on the spin-coupling to the magnetic field ${\bf B}$ generating the 
control flux, as exploited via Zeeman splitting in spin filters. It is a 
pure quantum interference effect due to the cooperation between change and spin
coherence during transport, which also exists for the non-averaged conductance at 
a given energy. We further find that this effect also pertains for large 
values of the Rashba strength $\delta_{\rm R}$ associated to wires of finite 
width $w$ ($Q_{\rm R}=(r_0/w)\delta_{\rm R}$), indicating that radial motion 
does not affect the control mechanism for spin-switching.
 
Additionally, further numerical calculations\cite{FRunp} for 2D ring 
structures supporting several open modes show features similar to that of
Fig.~\ref{spin-switch} for single-mode rings, as long
 as (i) the incoming and outgoing leads support just one open 
channel and (ii) the corresponding aspect ratio is small 
($w/r_0\ll 1$). Deviations from e.g. Fig.~\ref{spin-switch}(b) arise 
as $w/r_0$ increases, manifested by a less defined minimum in $G^{\ua \ua}$
at $\phi=\phi_0/2$ due to the relatively large fraction of flux $\phi$ 
penetrating the finite-width ring in that case. Moreover, asymmetric rings 
with arms of different effective length can also show a flux-modulated 
spin polarization similar to that of Fig.~\ref{spin-switch}(b). However, the 
spin switching is not complete, and it does not necessarily take place at 
$\phi=\phi_0/2$.


Analytical results for the spin-switching in 1D rings can be, in principle,
obtained by studying the spin resolved transmission probabitities 
$T^{\sigma'\sigma}$ as defined in Sec.~\ref{RMSUC}.

\section{conclusions}
\label{CONCL}

We have studied coherent spin-dependent transport in ballistic 
1D and 2D ring geometries subject to (spin-orbit) Rashba coupling. 
We first obtained, via analytical (1D) and numerical (2D) calculations, the 
spin-related conductance modulation profile of {\it unpolarized} spin 
carriers as a function of the scaled Rashba strength
$Q_{\rm R}$, which also acts as a measure defining adiabatic spin 
transport for $Q_{\rm R} \gg 1$. The conductance appears to be quite 
sensitive to $Q_{\rm R}$, suggesting the possibility of all-electrical 
spintronic devices. Moreover, we point out the role played by 
Aharonov-Anandan and Berry phases unnoticed in a previous 
proposal.\cite{NMT99}
In addition, we also studied the magneto conductance of 
{\it spin-polarized} carriers to assess possibilities for
controlling the spin orientation in the presence of Rashba 
coupling. We demonstrate that an additional small flux $\phi$ can be used as 
a control parameter for inducing spin flips. The mechanism arises from 
cooperative quantum interference of charge and 
spin degrees of freedom in coherent transport.
Combined with a spin detector such a device may be used for controlling
spin polarized currents alternative to the Datta-Das transistor.\cite{DD90}
Moreover, we note that the Dresselhaus spin-orbit coupling,\cite{D55} not studied here, 
could lead to similar conductance-modulation and spin-switching effects.
However, its interplay with the Rashba coupling in systems where both contributions are 
comparable can produce further effects of interest.\cite{SS03,SEL03}

\acknowledgements

We thank M. Governale, F. Meijer, J. Splettstoesser, and U. Z\"ulicke 
for useful discussions. We acknowledge financial support from the 
{\em Deutsche Forschungsgemeinschaft} and 
thank the {\it Max-Planck-Institut for the Physics 
of Complex Systems} in Dresden, Germany, for providing computational 
resources.

\vspace*{2mm}

\appendix

\section{1D spin eigenstates and effective Rashba field}
\label{App1}

The components of the eigenstates $\Psi_{\lambda,n}^s$ of the 1D Hamiltonian
(\ref{H1D}), which are given in Eq.~(\ref{spinor}) (spin $s=\pm 1$, 
travel direction $\lambda=\pm 1$, integer orbital number $n \ge 0$), 
satisfy the matrix equation
\bea
\label{H1Dspin}
\left( \begin{array}{cc}
\displaystyle{\frac{\hbar \omega_0}{2}} n'^2 + \frac{\hbar \omega_B}{2} &
\displaystyle{\frac{\hbar \omega_{\rm R}}{2}} \left(n'\!+\!\frac{1}{2}\right) \\
& \\
\displaystyle{\frac{\hbar \omega_{\rm R}}{2}}
\left(n'\!+\! \frac{1}{2}\right)& \displaystyle{\frac{\hbar \omega_0}{2}} (n'\!+\!1)^2
\! -\! \frac{\hbar \omega_B}{2}
\end{array}
\right) \chi = E_{\lambda,n}^s~\chi~~~
\eea
where the normalized spinors read
\bea
\label{spinorA2}
\chi=\left( \begin{array}{c}
\chi_1 \\
\chi_2 
\end{array}
\right)=\frac{1}{\sqrt{1+(\Delta_{\lambda,n}^s)^2}}
\left( \begin{array}{c}
1 \\
\Delta_{\lambda,n}^s 
\end{array}
\right),
\eea
with 
\bea
\Delta_{\lambda,n}^s=\frac{E_{\lambda,n}^s-(\hbar \omega_0/2)n'^2}
{(\hbar \omega_{\rm R}/2)(n'+1/2)}, 
\eea
$n'=\lambda n+\phi/\phi_0$, and eigenvalues given by
\begin{widetext}
\bea
\label{ES2}
E_{\lambda,n}^s&=&\frac{\hbar
\omega_0}{2}\left\{\left[\left(n'+\frac{1}{2}\right)^2\!+\!\frac{1}{4}\right]+
s \sqrt{\left[\left(n'+\frac{1}{2}\right)- \frac{\omega_B}{\omega_0}\right]^2
+\left(\frac{\omega_{\rm R}}{\omega_0}\right)^2 \left(n'+\frac{1}{2}\right)^2}
\right\} .
\eea
\end{widetext}
The off-diagonal elements on the left-hand side of Eq.~(\ref{H1Dspin}) 
determine the magnitude and orientation of the in-plane effective Rashba field 
${\bf B}_{\rm R}$. The resulting overall effective field 
${\bf B}_{\rm eff}={\bf B}+
{\bf B}_{\rm R}$ has a tilt angle $\alpha$ with respect to the $z$-axis
satisfying $\tan \alpha = \omega_{\rm R} (n'+1/2)/\omega_B$. Moreover, the
presence of the kinetic terms in the diagonal elements of (\ref{H1Dspin}) 
prevent the spinors $\chi$
to align with ${\bf B}_{\rm eff}$. Instead, they are characterized by a
tilt angle
$\gamma$ which tends to $\alpha$ only for strong spin coupling (adiabatic 
limit). 

For illustration we discuss the spin-up case ($s=1$) and the
dependence on the travel direction $\lambda$ in the absence of Zeeman coupling 
($\omega_B=0$, $\alpha=\pi/2$) provided that a finite flux $\phi$ in present. 
Then we find from (\ref{spinorA2})
\bea
\label{chi1}
\chi_1^\ua&=& \frac{Q_{\rm R}}{\sqrt{2}\left[Q_{n'} +Q_{\rm R}^2\right]^{1/2} } \\
\label{chi2}
\chi_2^\ua&=&\frac{Q_{n'}}{\sqrt{2}\left[Q_{n'} +Q_{\rm R}^2\right]^{1/2} } \; ,
\eea
where $Q_{n'} = 1+{\rm sign}[n'+1/2]\sqrt{1+Q_{\rm R}^2}$.
The dimensionless Rashba strength $Q_{\rm R}$ is defined in Eq.~(\ref{eq:QR}).
In the adiabatic, strong coupling limit ($Q_{\rm R} \gg 1$)
we obtain from (\ref{chi1}) and (\ref{chi2})
\bea
\label{chiup1}
\chi^\ua \stackrel{Q_{\rm R}\gg 1}{\longrightarrow} 
\left\{ \begin{array}{c}
\left(
\begin{array}{c}
1/\sqrt{2}^-\\
1/\sqrt{2}^+
\end{array}
\right)~~{\rm if~sign}[n'+1/2]=1~\\
\\
\left(
\begin{array}{c}
1/\sqrt{2}^+\\
-1/\sqrt{2}^+
\end{array}
\right)~~{\rm if~sign}[n'+1/2]=-1\\
\end{array}
\right.
\eea
indicating that the spinors are contained within the plane defined by the 
ring and pointing along ${\bf B}_{\rm R}$. On the other hand, in the 
opposite limit of weak coupling ($Q_{\rm R}\ll 1$), we arrive at  
\bea
\label{chiup2}
\chi^\ua \stackrel{Q_{\rm R}\ll 1}{\longrightarrow} 
\left\{ \begin{array}{c}
\left(
\begin{array}{c}
0^+\\
1^-
\end{array}
\right)~~{\rm if~sign}[n'+1/2]=1~\\
\\
\left(
\begin{array}{c}
1^-\\
0^-
\end{array}
\right)~~{\rm if~sign}[n'+1/2]=-1\\
\end{array}
\right.
\eea
highlighting the influence of the travelling direction on
the relative orientation of the spinors. As a consequence we find that the up 
spinors can be written as
\bea
\label{chiup3}
\chi^\ua = 
\left\{ \begin{array}{c}
\left(
\begin{array}{c}
\sin \gamma/2\\
\cos \gamma/2
\end{array}
\right)~~{\rm if~sign}[n'+1/2]=1~\\
\\
\left(
\begin{array}{c}
\cos \gamma/2\\
-\sin \gamma/2
\end{array}
\right)~~{\rm if~sign}[n'+1/2]=-1\\
\end{array}
\right.
\eea
with $\tan \gamma =Q_{\rm R}$. Following a similar procedure for $\phi=0$ 
($n'=\lambda n$) we find the eigenstates listed in 
Sec.~\ref{1Deigenstates}.

%
%


\begin{thebibliography}{50}
\bibitem{note14} We neglect the Dresselhaus spin-orbit coupling,\cite{D55}
having in mind e.g. InAs or InSb semiconductors where the Rashba interaction dominates.\cite{SS03}
\bibitem{note1} The correction (see Ref.~\onlinecite{MMK02}) to the Hamiltonian 
used in earlier work consists in adding the last term in Eq.~(\ref{H1D}), 
which becomes negligible for large angular momentum 
($\langle -i \partial/\partial \varphi + \phi/\phi_0 \rangle \gg 1$).
\bibitem{note4} This make sense since we see from Sec.~\ref{1Deigenstates} 
that such pairs of spins are actually parallel.
\bibitem{note5} Note that the phase $\pi Q_{\rm R}$ is equivalent to the 
differential Rashba phase $\Delta \theta=\alpha_{\rm R}(2m^*/\hbar^2)L$ 
acquired in a straight wire of length $L=\pi r_0$.\cite{DD90,MK01}
\bibitem{note7} The number of incoming and outgoing open channels is kept constant within such 
energy window. Physically, the energy-average can by associated either with an ensemble-average 
on rings of slightly different size or, alternatively, with a finite temperature (sufficiently low 
for neglecting decoherence effects).
\bibitem{note12} Note that here the term {\it adiabatic} is used in a sense different from 
the one introduced previously in the context of adiabatic spin transport.
\bibitem{note13} The adiabatic switching of the Rashba coupling is introduced in order to avoid 
an abrupt transition between the region of interest subject to finite coupling (the ring) and 
the leads connected to reservoirs free of spin-orbit coupling. For a given energy, some 
differences can arise in the computed conductance when using or not the adiabatic switching in 
the case of strong Rashba coupling. However, no significant differences survive when performing 
an energy average in the way discribed above.
\end{thebibliography}



\end{document}